\documentclass[aps,prd,twocolumn,citeautoscript,nofootinbib,showkeys,superscriptaddress]{revtex4-1}                  
\pdfoutput=1 
\usepackage{amsmath,amssymb,mathrsfs,bm,feynmf,setspace}
\usepackage{amsthm}
\usepackage{graphicx}     
\usepackage[tight]{subfigure}     
\usepackage[usenames, dvipsnames]{color} 
\usepackage{braket} 
\usepackage{tensor} 
\usepackage[dvipsnames]{xcolor}
\definecolor{cadmiumred}{rgb}{0.89, 0.0, 0.13}
\definecolor{darkblue}{rgb}{0.2, 0, 0.8}
\definecolor{darkgreen}{rgb}{0.2, 0.71, 0}  
\definecolor{LinkGreen}{cmyk}{1,0.05,.95,0.5}
\usepackage[colorlinks=true]{hyperref}  
\hypersetup{
    bookmarks=true,         
    unicode=false,          
    pdftoolbar=true,        
    pdfmenubar=true,        
    pdffitwindow=false,     
    pdfstartview={FitH},    
    pdftitle={My title},    
    pdfauthor={Author},     
    pdfsubject={Subject},   
    pdfcreator={Creator},   
    pdfproducer={Producer}, 
    pdfkeywords={keyword1} {key2} {key3}, 
    pdfnewwindow=true,      
    colorlinks=true,       
    linkcolor=Blue, 
    citecolor=cadmiumred,        
    filecolor=blue,      
    urlcolor=Blue         
} 

\newtheorem*{hypothesis*}{Hypothesis}

\newcommand{\vev}[1]{\langle #1 \rangle}

\newcommand{\req}[1]{(\ref{#1})} 

\newcommand{\bea}{\begin{eqnarray}}
\newcommand{\eea}{\end{eqnarray}}
\newcommand{\ba}{\begin{eqnarray}}
\newcommand{\ea}{\end{eqnarray}}

\newcommand{\beq}{\begin{equation}}
\newcommand{\eeq}{\end{equation} }
\newcommand{\beqa}{\begin{eqnarray}}
\newcommand{\eeqa}{\end{eqnarray}}
\newcommand{\beqar}{\begin{eqnarray*}}
\newcommand{\be}{\begin{equation}}
\newcommand{\ee}{\end{equation}}
\newcommand{\eeqar}{\end{eqnarray*}}

\newcommand{\ssc}{\scriptscriptstyle}

\newcommand{\see}{S_{\rm \ssc EE}}

\newcommand{\dd}{\mathrm{d}}
\newcommand{\lie}{\pounds}

\begin{document}

\title{Entanglement equilibrium for  higher order gravity}
\author{Pablo Bueno}
\email{pablo@itf.fys.kuleuven.be}
\affiliation{Instituut voor Theoretische Fysica, KU Leuven, 
Celestijnenlaan 200D, B-3001 Leuven, Belgium}
\affiliation{Institute for Theoretical Physics, University of Amsterdam, 1090 GL Amsterdam, The
Netherlands}

\author{Vincent S. Min}
\email{vincent.min@kuleuven.be}
\affiliation{Instituut voor Theoretische Fysica, KU Leuven, 
Celestijnenlaan 200D, B-3001 Leuven, Belgium}

\author{Antony J. Speranza}
\email{asperanz@gmail.com}
\affiliation{Maryland Center for Fundamental Physics, University of Maryland, College Park, MD
20742, USA}

\author{Manus R. Visser}
\email{m.r.visser@uva.nl}
\affiliation{Institute for Theoretical Physics, University of Amsterdam, 1090 GL Amsterdam, The
Netherlands}
\date{\today}
\begin{abstract}     
We show that the linearized higher derivative gravitational field equations are equivalent to 
an equilibrium condition on the entanglement entropy of small spherical regions in vacuum.  This 
extends Jacobson's   recent derivation of the Einstein equation using entanglement to 
include general higher derivative corrections.  The corrections are naturally associated with the 
subleading divergences in the entanglement entropy, which take the form of a Wald
entropy evaluated on the entangling surface.
Variations of this Wald entropy are related to the field equations through an
identity for causal diamonds in maximally symmetric spacetimes, which we derive for 
arbitrary higher derivative theories.
If the variations are taken holding fixed a geometric functional that we call the generalized
volume, the identity becomes an equivalence between the linearized constraints and the 
entanglement equilibrium condition.
We note that the fully nonlinear higher curvature equations cannot be derived from the 
linearized equations applied to small balls, in contrast to the situation encountered
in Einstein gravity.  The generalized volume is a novel result of this work, and we
speculate on its thermodynamic role in the first law of causal diamond mechanics, as well
as its possible application to holographic complexity.  
\end{abstract}   
\maketitle


\section{Introduction} \label{sec:intro}

Black hole entropy 
remains one of the best windows into the nature 
of quantum gravity available to dwellers of the infrared. 
Bekenstein's original motivation for introducing it
was to avoid gross violations of the 
second law of thermodynamics by sending matter
into the 
black hole, decreasing the entropy of the exterior   
\cite{Bekenstein1972, Bekenstein1973a}.  
The subsequent discovery by Hawking that black holes radiate thermally at 
a temperature $T=\kappa/2\pi$, with $\kappa$ the surface gravity,
fixed the value of the entropy in terms of the area to be
$
S_\text{BH} = A/4G,
$
and suggested  a deep connection to  quantum properties
of gravity \cite{Hawking1974}.

The appearance of area in $S_\text{BH}$ is somewhat mysterious from
a classical perspective; however, an intriguing explanation emerges by considering 
the entanglement entropy of quantum fields outside the horizon \cite{Sorkin:2014kta,
Bombelli1986, Srednicki1993a, Frolov1993a}.  Entanglement entropy is UV divergent,
and upon regulation it takes the form
\beq \label{eqn:SEE}
S_\text{EE} = c_0 \frac{A}{\epsilon^{d-2}} +\{\text{subleading divergences}\} + S_\text{finite} \, ,
\eeq
with $\epsilon$ a regulator and $c_0$ a constant. 
Identifying the coefficient $c_0/\epsilon^{d-2}$ with $1/4G$ would allow $S_\text{BH}$
 to be attributed to the leading divergence in the entanglement entropy. 
The subleading divergences 
could similarly be associated with higher curvature gravitational couplings,  
 which change the expression for the black hole entropy 
to  the Wald entropy \cite{Wald1993}.

To motivate these identifications, one must assume that the quantum gravity theory
is UV finite (as occurs in string theory), yielding a finite 
entanglement entropy,  cut off 
near the Planck length, $\epsilon \sim \ell_P$.  
Implementing this cutoff would seem to depend on a detailed knowledge  
of the UV theory, 
inaccessible from the vantage of low energy effective field theory.  Interestingly, this issue can be resolved
 within the effective theory by the renormalization of  the gravitational
couplings by matter loop divergences.  
There is mounting evidence that 
these precisely match the entanglement entropy divergences, making 
the {\it generalized entropy}
\beq \label{eqn:Sgen}
S_\text{gen} = S_\text{Wald}^{(\epsilon)}+ S^{(\epsilon)}_\text{mat}
\eeq  
independent of $\epsilon$ \cite{Susskind1994, 
Cooperman2013, Solodukhin2011a, Bousso2016}. Here $S_\text{Wald}^{(\epsilon)}$ 
is the Wald entropy expressed in terms of the renormalized gravitational couplings and
$S^{(\epsilon)}_\text{mat}$ is a renormalized 
entanglement entropy of matter fields that is related to $S_\text{finite}$ in 
(\ref{eqn:SEE}), although the precise relation depends on the 
renormalization scheme.\footnote{A covariant 
regulator must be used to ensure that the subleading
divergences appear as  a Wald entropy.  Also, since power law divergences are not universal, 
when they are present the same renormalization scheme must be used for the 
   entanglement entropy
and the gravitational couplings.  Additional subtleties for nonminimally coupled fields, gauge
fields, and gravitons are discussed in Section \ref{sec:nonmin}. }  
The identification of gravitational couplings with entanglement entropy divergences is
 therefore consistent with the renormalization group (RG)
flow in the low energy effective theory, and amounts to assuming that the bare
gravitational couplings vanish \cite{Jacobson1994a}.  In this picture, 
$S_\text{gen} = S_\text{EE}$, with $S_\text{Wald}^{(\epsilon)}$ acting
as a placeholder for the UV degrees of freedom that have been integrated out.

When viewed as entanglement entropy, 
it is clear that generalized entropy can be assigned to surfaces other than black hole
horizon cross sections 
\cite{Jacobson1999, Jacobson2003a, 
Bianchi2012a, Bousso2016}.  For example, in holography
the generalized entropy of a minimal surface in the bulk is dual via the quantum-corrected
Ryu-Takayanagi formula \cite{Ryu:2006bv, Faulkner2013a} to the entanglement
entropy of a region of the boundary CFT.\footnote{The 
UV divergences in the CFT entanglement entropy have no relation to the Planck length
in the bulk, but instead are related to the infinite area of the minimal surface in AdS, courtesy
of the UV/IR correspondence.}
Even without assuming  holographic duality, 
the generalized entropy  provides a link between the geometry of  surfaces and  
entanglement entropy.   When supplemented with thermodynamic 
information, this link can give rise to dynamical equations for gravity.  
The first demonstration of this was Jacobson's derivation of the Einstein equation as an 
equation of state for local causal horizons possessing an entropy proportional to their area 
\cite{Jacobson1995a}.  Subsequent work using entropic arguments 
\cite{Verlinde2010,Verlinde2016} and holographic entanglement entropy \cite{Lashkari2013, Faulkner2013,Swingle2014a} confirmed that entanglement thermodynamics is connected to 
gravitational dynamics.

\begin{figure}[tp!]
\includegraphics[width=\columnwidth]{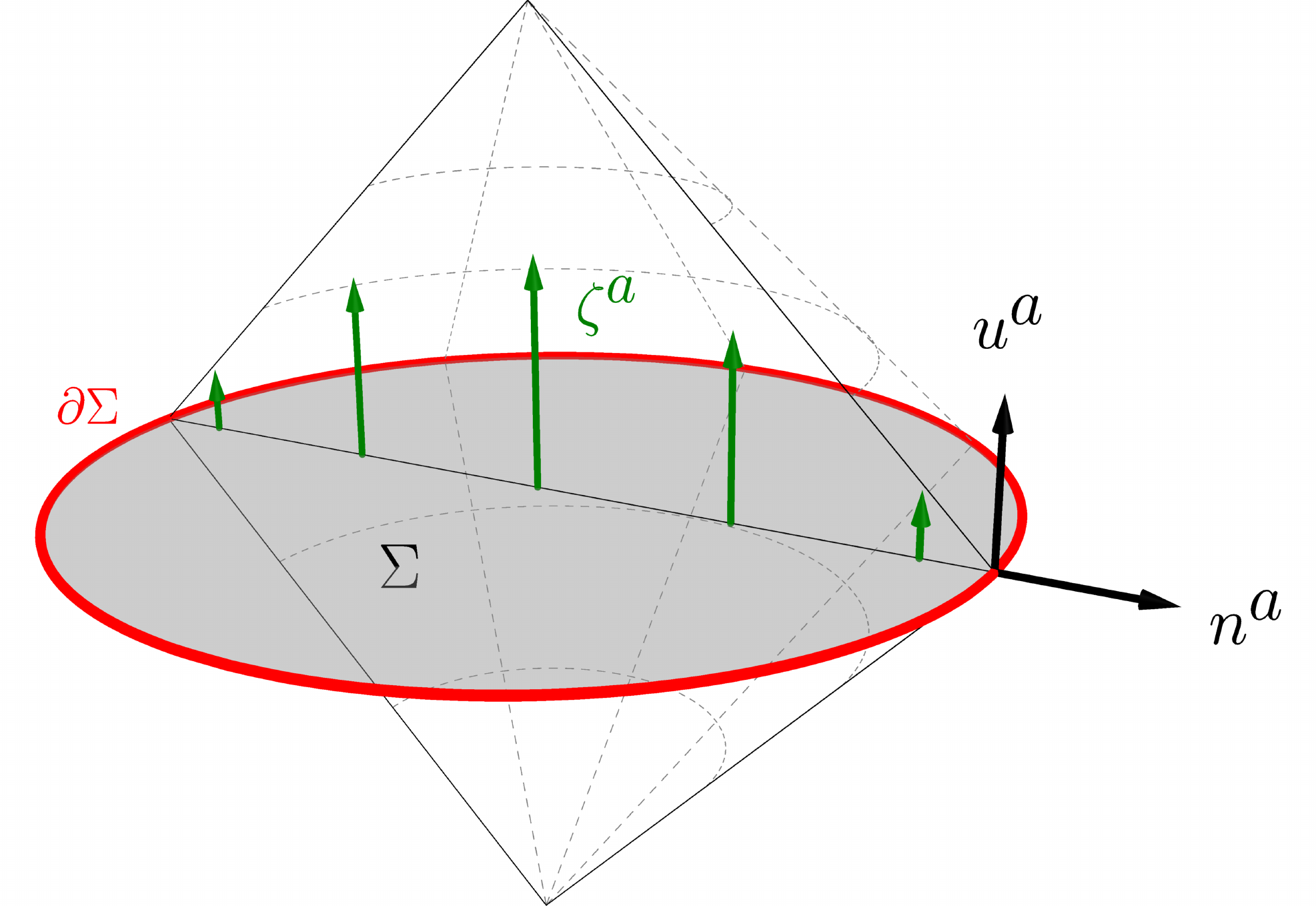}
\caption{The causal diamond consists of the future and past domains
of dependence of a spatial sphere $\Sigma$ in a MSS. 
$\Sigma$ has a unit normal $u^a$, induced metric $h_{ab}$ and volume form $\eta$.  
The boundary  $\partial\Sigma$ has a spacelike unit normal $n^a$ defined
to be orthogonal to $u^a$,  and volume form $\mu$.  
The conformal Killing vector $\zeta^a$ generates a flow within the causal diamond, and 
 vanishes on  the bifurcation surface $\partial\Sigma$. \label{fig:diamond}}
\end{figure}

Recently, Jacobson has advanced a new viewpoint on the relation between geometry 
and entanglement that has been dubbed ``entanglement equilibrium'' \cite{Jacobson:2016aa}.
This proposal considers spherical spatial subregions in geometries that are a 
perturbation of a maximally symmetric spacetime (MSS).  Each such subregion $\Sigma$
 in the maximally
symmetric background defines a causal diamond, which admits a conformal Killing vector 
$\zeta^a$
whose flow preserves the diamond (see Figure \ref{fig:diamond}).  
The entanglement equilibrium hypothesis states that any perturbation of the matter
fields and geometry inside the ball leads to a decrease in entanglement, i.e., the vacuum
is a maximal  entropy state.  
This hypothesis applies holding the volume of $\Sigma$ fixed; even so, 
the
introduction of curvature from the geometry variation 
can lead to a decrease in the area of the boundary $\partial\Sigma$. 
This affects the divergent terms in the entanglement entropy by changing Wald entropy,
which at leading order is simply $A/4G$. The variation of the quantum
state  contributes a piece $\delta S_\text{mat}$, and maximality implies the   total
variation of the entanglement entropy vanishes at first order,\footnote{The separation
of the entanglement entropy into a divergent Wald piece and a finite matter piece is 
scheme-dependent, and can change under the RG flow \cite{Jacobson:2012ek}.  
Also the matter variation can sometimes produce state-dependent divergences
\cite{Marolf2016}, which appear as a variation of the Wald entropy.  Since
we only ever deal with total variations of the generalized entropy, these subtleties do not affect 
any results.  For simplicity, we will refer to $\delta S_\text{Wald}$ as coming from
the geometry variation, and $\delta S_\text{mat}$ from the matter state variation. } 
\beq \label{eqn:dSEEV}
\delta S_\text{EE}\big|_V = \frac{\delta A\big|_V}{4G} + \delta S_\text{mat} = 0 \, .
\eeq
When applied to small spheres, this maximal entropy 
condition was 
shown to be equivalent to imposing the Einstein equation at the center of the ball.

Taken as an effective field theory, gravity is expected to contain higher curvature corrections 
that arise from matching to its UV completion.  An important test 
of the entanglement equilibrium hypothesis is whether it can consistently accommodate these 
corrections.  It is the purpose of this paper to demonstrate that a generalization
to higher curvature theories is possible 
 and relates to  
the subleading divergences appearing in (\ref{eqn:SEE}).

\subsection{Summary of results and outline}
 It is not {\it a priori} clear what the precise statement of the entanglement equilibrium condition
should be for a higher curvature theory, and in particular what replaces the fixed-volume
constraint. 
The  formulation we propose here  is advised by
the {\it first law of causal diamond mechanics}, a purely geometrical identity that holds independently of 
any entanglement considerations.  
It was derived  for Einstein gravity
in the supplemental materials of \cite{Jacobson:2016aa}, and one of the 
main results of this paper is to extend it to arbitrary higher derivative theories.  
As we show in section \ref{sec:firstlaw}, the first law is related to the off-shell identity 
\beq\label{titis}
\frac{\kappa}{2\pi} \delta S_{\text{Wald}}\big|_W + \delta H^m_\zeta = \int_\Sigma \delta C_\zeta \, ,
\eeq   
where $\kappa$ is the surface gravity of $\zeta^a$ \cite{Jacobson1993}, $S_\text{Wald}$ is the 
Wald entropy of $\partial \Sigma$ given in equation (\ref{eqn:SWald}) \cite{Wald1993, Iyer1994a},
$H_\zeta^m$ is the matter Hamiltonian for flows along $\zeta^a$, defined in equation 
(\ref{potati}), and $\delta C_\zeta=0$ are the linearized constraint equations of the higher
 derivative  theory.  The Wald entropy is varied holding fixed a local geometric 
functional
\beq \label{eqn:Wolume}
W = \frac{1}{(d-2)E_0} \int_\Sigma {\eta}\left(E^{abcd} u_a h_{bc} u_d - E_0\right) \, ,
\eeq
with $\eta$, $u^a$ and $h_{ab}$ defined in Figure \ref{fig:diamond}.  
$E^{abcd}$ is the variation of the gravitational Lagrangian scalar with respect to 
$R_{abcd}$ and $E_0$ is a constant determined by the value of $E^{abcd}$ in a MSS via
$E^{abcd}\overset{\text{MSS}}{=}E_0(g^{ac}g^{bd}-g^{ad}g^{bc})$.
We refer to $W$ as the ``generalized volume'' since it reduces to the volume for
Einstein gravity.

The Wald formalism contains ambiguities identified by Jacobson, Kang and Myers (JKM)
\cite{Jacobson1994b}  that modify the  Wald entropy and the 
generalized volume by the terms $S_\text{JKM}$ and $W_\text{JKM}$ 
given in (\ref{eqn:SJKM}) and (\ref{eqn:WJKM}).  Using a modified generalized
volume defined by
\beq \label{eqn:Wp}
W' = W + W_\text{JKM} \, ,
\eeq
the identity (\ref{titis}) continues to hold with $\delta (S_\text{Wald}+S_\text{JKM})\big|_{W'}$
replacing $\delta S_\text{Wald}\big|_W$\,.  As discussed in section \ref{sec:subleading}, the subleading divergences for the 
entanglement entropy involve a particular resolution of the JKM ambiguity, while 
section \ref{sec:fixedflux} argues that
the first law of causal diamond mechanics applies for \emph{any} resolution, as long as the 
appropriate generalized volume is held fixed.

Using the resolution of the JKM ambiguity required for the entanglement entropy calculation,
the first law leads to the following statement of entanglement
equilibrium, applicable to higher curvature theories:
\begin{hypothesis*}[Entanglement Equilibrium]
In a quantum gravitational theory, the entanglement entropy of a spherical 
region with fixed generalized volume $W'$ is maximal in vacuum.
\end{hypothesis*}
This modifies the original equilibrium condition (\ref{eqn:dSEEV}) by replacing the area
variation with 
\beq
\delta(S_\text{Wald} + S_\text{JKM})\big|_{W'} \, .
\eeq
In Section \ref{sec:equilibrium}, this equilibrium condition is shown
to be equivalent to   the linearized higher derivative field equations 
in the case that the matter fields are conformally invariant.\footnote{There is a proposal
for including nonconformal matter that involves varying a local cosmological constant
\cite{Jacobson:2016aa, Casini2016, Speranza2016}.  If valid, that proposal applies 
in the higher curvature case as well, since it deals only with the matter variations.}  
Facts about  
entanglement entropy divergences and 
the reduced density matrix for a sphere in a CFT are used to relate the total variation of 
the entanglement entropy to the left hand 
side of (\ref{titis}).  Once this is done, it becomes clear that imposing the linearized 
constraint equations is 
equivalent to the entanglement equilibrium condition.

In \cite{Jacobson:2016aa}, this condition was applied in the small
ball limit, in which {\it any} geometry looks like a perturbation of a MSS.  Using Riemann
normal coordinates (RNC), the linearized equations were shown to impose the fully nonlinear
equations for the case of Einstein gravity.  
We will discuss this argument in Section \ref{sec:equations} for higher
curvature theories, and show that the nonlinear equations can \emph{not} be obtained from the 
small ball limit, making general relativity unique in that regard.

In section \ref{sec:conclusion}, we discuss several implications of this work.  First, we describe
how it compares to other approaches  connecting geometry and entanglement.  
Following that, we provide a possible thermodynamic interpretation of the first law of causal 
diamond mechanics derived in section \ref{sec:firstlaw}.  We then comment on a conjectural
relation between our generalized volume $W$ and higher curvature holographic complexity.  
Finally, we lay out several future directions for the entanglement equilibrium program.

Note on conventions: we set $\hbar=c=1$, use metric signature $(-,+,+,\ldots)$, and 
use $d$ to refer to the spacetime dimension.   
We write the spacetime volume form as $\epsilon$, and 
occasionally we will denote it $\epsilon_a$ or $\epsilon_{ab}$, suppressing all but its first one or 
two abstract indices.

\section{First law of  causal diamond mechanics} \label{sec:firstlaw}

Jacobson's 
entanglement equilibrium argument \cite{Jacobson:2016aa} 
compares the surface area of a 
small spatial ball $\Sigma$ in a curved spacetime to the one that would be obtained in a MSS.  
The comparison is made using balls of equal volume $V$, a choice justified 
by an Iyer-Wald variational identity \cite{Iyer1994a} 
for the conformal Killing vector $\zeta^a$ of the causal diamond in the maximally symmetric background. When the Einstein equation holds, this identity implies the 
{\it first law of causal diamond mechanics} \cite{Jacobson:2016aa, Manus}
\beq \label{barbecue}
- \delta H_\zeta^m = \frac{\kappa}{8\pi G} \delta A   -\frac{\kappa k}{8\pi G}  \delta V \, ,
\eeq
where $k$ is 
  the trace of the extrinsic curvature of $\partial \Sigma$ embedded in $\Sigma$,
and the matter conformal Killing energy $H_\zeta^m$ is constructed from the  stress 
tensor $T_{ab}$ by
\begin{equation}\label{potati}
 H_{\zeta}^{m}=\int_\Sigma\eta\, u^a \zeta^bT_{ab}\, .
 \end{equation}
 The purpose of this section is to generalize the variational identity to higher derivative
 theories, and to clarify its relation to the equations of motion.  
This is done by focusing on an off-shell version of the identity, which reduces to the first law
when the  linearized  constraint equations for the theory are satisfied.  We begin by reviewing the 
Iyer-Wald formalism in subsection \ref{sec:IyerWald}, which also serves to establish notation.
After describing the geometric setup in subsection \ref{subsec:setup}, we show in subsection
\ref{sec:localgeo} how the quantities appearing in the identity can be written as variations of 
local geometric functionals of the surface $\Sigma$ and its boundary $\partial \Sigma$.  As
one might expect, the area is upgraded to the Wald entropy $S_\text{Wald}$, and we derive 
the generalization of the volume  given in equation (\ref{eqn:Wolume}). 
Subsection \ref{sec:fixedflux} describes how the variational identity can instead be viewed 
as a variation at fixed generalized volume $W$, as quoted in equation (\ref{titis}), 
and describes the effect that JKM
ambiguities have on the setup.

\subsection{Iyer-Wald formalism} \label{sec:IyerWald}
We begin by recalling the Iyer-Wald formalism \cite{Wald1993, Iyer1994a}.  A general diffeomorphism invariant theory may 
be defined by its Lagrangian $L[\phi]$, a spacetime $d$-form locally constructed 
from the dynamical fields $\phi$, which include the metric and matter fields.  A variation of this
Lagrangian takes the form  
\beq \label{eqn:dL}
\delta L = E\cdot  \delta\phi + \dd\theta[\delta\phi] \, ,
\eeq
where  $E$ collectively denotes the equations of motion for the dynamical fields, and $\theta$
is the symplectic potential $(d-1)$-form.  Taking an antisymmetric variation of $\theta$ yields
the symplectic current $(d-1)$-form
\beq \label{eqn:symplectomega}
\omega[\delta_1\phi, \delta_2\phi] = \delta_1 \theta[\delta_2\phi] - \delta_2\theta[\delta_1\phi] \,,
\eeq
whose integral over a Cauchy surface $\Sigma$ gives the symplectic form for the phase space description
of the theory.  Given an \emph{arbitrary} vector field $\zeta^a$, 
   evaluating the symplectic
form on the Lie derivative  $\lie_\zeta \phi$  
gives the variation of the Hamiltonian $H_\zeta$ that generates the flow of  $\zeta^a$ 
\beq \label{eqn:hamilton}
\delta H_\zeta = \int_\Sigma \omega[\delta\phi, \lie_\zeta\phi] \,.
\eeq
Now consider   a ball-shaped region $\Sigma$, and take $\zeta^a$ to be any future-pointed,
timelike vector that vanishes on the boundary $\partial\Sigma$. Wald's variational identity then
 reads 
\beq \label{eqn:dHz}
\int_\Sigma \omega[\delta\phi, \lie_\zeta\phi] =  \int_\Sigma \delta J_\zeta \,,
\eeq
where the Noether current $J_\zeta$ is defined by
\beq
J_\zeta = \theta[\lie_\zeta \phi]-i_\zeta L \,.
\eeq
Here $i_\zeta$ denotes contraction of the vector $\zeta^a$ on the first index of the differential
form $L$. The identity (\ref{eqn:dHz}) holds when the background geometry satisfies
the field equations $E=0$, and it assumes that $\zeta^a$ vanishes on $\partial \Sigma$.  
Next we note that the 
Noether current can always be expressed as \cite{Iyer1995a}
\beq  \label{eqn:noethercurrent}
J_\zeta = \dd Q_\zeta + C_\zeta,
\eeq
where $Q_\zeta$ is the Noether charge $(d-2)$-form and $C_\zeta$ are the constraint
field equations, which arise as a consequence of the diffeomorphism gauge symmetry. 
For non-scalar matter, these constraints are a combination of the metric and matter field
equations \cite{Seifert2007b, Jacobson2011a}, but, 
assuming the matter equations  are imposed, 
we can take $C_\zeta = -2 \zeta^a E\indices{_a^b}\epsilon_b$, where $E^{ab}$ is the variation
of the Lagrangian density with respect to the  metric.  
By combining equations (\ref{eqn:hamilton}), (\ref{eqn:dHz}) and (\ref{eqn:noethercurrent}), one finds that 
\beq \label{eqn:FLDM}
-\int_{\partial\Sigma} \delta Q_\zeta+\delta H_\zeta =   \int_\Sigma \delta C_\zeta \, .
\eeq
When the linearized constraints hold, $\delta C_\zeta = 0$, the variation of the Hamiltonian
is a boundary integral of $\delta Q_\zeta$.  This on-shell identity forms the basis for 
deriving the first law of causal diamond mechanics.  Unlike the situation encountered
in black hole thermodynamics, $\delta H_\zeta$ is not zero because below we take $\zeta^a$ to be a conformal Killing vector as opposed to a true Killing vector.

\subsection{Geometric setup}
\label{subsec:setup}

Thus far, the only restriction that has been placed on the vector field $\zeta^a$ is that it 
vanishes on $\partial\Sigma$.  
As such, the quantities $\delta H_\zeta$ and $\delta Q_\zeta$ appearing 
in the identities depend rather explicitly on the fixed vector $\zeta^a$, and therefore these
quantities are not written in terms of only the geometric properties of the surfaces
$\Sigma$ and $\partial\Sigma$.  A purely geometric description is desirable if
the Hamiltonian and Noether charge are 
to be interpreted as  
thermodynamic state functions, which ultimately may be used to define the 
ensemble of geometries in any proposed quantum description of the microstates.  
This situation may be remedied by choosing the vector $\zeta^a$ and the surface
$\Sigma$ to have special properties in the background geometry.  In particular, by choosing
$\zeta^a$ to be a conformal Killing vector for a causal diamond in the MSS, and picking
$\Sigma$ to be the surface on which the conformal factor vanishes, one finds that 
the perturbations $\delta H_\zeta$ and $\delta Q_\zeta$    have expressions in terms of local
geometric functionals on the surfaces  $\Sigma$ and $\partial\Sigma$, respectively.

Given a causal diamond in a MSS,
there exists a conformal Killing vector $\zeta^a$ which generates a flow within the diamond and 
vanishes at the bifurcation surface $\partial\Sigma$ (see figure \ref{fig:diamond}).  
The metric satisfies the conformal Killing equation
\beq  \label{eqn:lie}
\lie_\zeta g_{ab} =    2\alpha g_{ab} \quad \text{with} \quad \alpha = \frac{1}{d}  \nabla_c \zeta^c  \,.
\eeq
and the conformal factor $\alpha$ vanishes on the spatial ball $\Sigma$.  
The gradient of $\alpha$ is hence proportional to the unit normal to $\Sigma$, 
\beq
u_a = N \nabla_a \alpha  \quad \text{with} \quad N =  \lVert \nabla_a \alpha \rVert^{-1} .
\eeq
Note the vector $u^a$ is future pointing since the conformal factor $\alpha$ decreases to the 
future of $\Sigma$. 
In a MSS, the normalization 
function $N$ has the curious property that it is constant over   $\Sigma$, and is
given by \cite{Manus}
\beq \label{eqn:N}
N =   \frac{d-2}{\kappa  k},
\eeq
where $k$ is the trace of the extrinsic curvature of $\partial \Sigma$ embedded in $\Sigma$, and $\kappa$ is the surface gravity of the conformal Killing horizon, defined momentarily.  
This constancy ends
up being crucial to finding a local geometric functional for $\delta H_\zeta$. 
Throughout this work, $N$ and $k$ will respectively
denote  constants equal to the normalization
function and extrinsic curvature trace, both evaluated in the background spacetime.  

Since $\alpha$ vanishes on $\Sigma$, $\zeta^a$ is instantaneously a Killing vector.  On the other hand, the covariant derivative of $\alpha$  is nonzero, so
\beq \label{eqn:covlie}
\nabla_d (\lie_\zeta g_{ab}) \big |_\Sigma = \frac{2}{N} u_d g_{ab} \, .
\eeq
The fact that the covariant derivative is nonzero on  $\Sigma$
is responsible for making   $\delta H_\zeta$ nonvanishing.

A conformal Killing vector with a horizon has a well-defined surface gravity 
$\kappa$ \cite{Jacobson1993},
and since $\alpha$ vanishes on $\partial\Sigma$, we can conclude that 
\beq\label{eqn:dz}
\nabla_a \zeta_b  \big  |_{\partial \Sigma}= \kappa n_{ab}\, , 
\eeq
where $n_{ab} = 2u_{[a} n_{b]}$ is the binormal for the surface $\partial\Sigma$,  
and $n^b$ is the outward
pointing spacelike unit normal to $\partial\Sigma$.  
Since $\partial\Sigma$ is a bifurcation surface of a conformal Killing horizon, 
$\kappa$  is  constant everywhere on 
it.
We provide an example of these constructions in appendix \ref{appkill} where
we discuss the conformal Killing vector for a causal diamond in flat space.

\subsection{Local geometric expressions} \label{sec:localgeo}

In this subsection we    evaluate the Iyer-Wald identity (\ref{eqn:FLDM})   for an arbitrary higher derivative theory of gravity and for the geometric setup described above. The final on-shell 
result is given in (\ref{firstlawhigher}), which is the first law of causal diamond mechanics for higher derivative gravity.

Throughout the computation we   assume that the  matter fields are minimally coupled, so that the Lagrangian splits into a metric and matter piece $L = L^g+L^m$, 
 and we take $L^g$ to be an \emph{arbitrary}, 
diffeomorphism-invariant function of the metric, Riemann tensor, and 
its covariant derivatives.  The symplectic potential and variation of the Hamiltonian  then   exhibit a similar  separation, $\theta  = \theta^g +\theta^m$ and $\delta H_\zeta = \delta H_\zeta^g+ \delta H_\zeta^m$, and so we can write equation (\ref{eqn:FLDM}) as 
\beq \label{newvarid}
-\int_{\partial \Sigma} \delta Q_\zeta+ \delta H_\zeta^g + \delta H_\zeta^m =  \int_\Sigma \delta C_\zeta \,.
\eeq
Below, we explicitly    compute the two terms $\delta H_\zeta^g$ and $\int_{\partial \Sigma} \delta Q_\zeta$ 
for the present geometric context. \\

 \paragraph{Wald entropy}

By virtue of equation (\ref{eqn:dz})  and the fact that $\zeta^a$ vanishes on $\partial\Sigma$, one can show that the integrated Noether charge
is simply related to the Wald entropy \cite{Wald1993, Iyer1994a}
\begin{align}
 -\int_{\partial\Sigma} 
Q_\zeta &=   \int_{\partial\Sigma}   \, E^{abcd}  \, \epsilon_{ab} \nabla_c \zeta_d \nonumber\\
&=\frac{\kappa}{2\pi}S_\text{Wald} \,, \label{eqn:Sbar}
\end{align}
where the Wald entropy is defined as
\beq \label{eqn:SWald}
S_\text{Wald} =  -  2\pi \int_{\partial\Sigma} \mu  \, E^{abcd} n_{ab} n_{cd} \,.
\eeq
 $E^{abcd}$ is the variation of the Lagrangian scalar   with respect to the 
Riemann tensor $R_{abcd}$  taken as an independent field, given in (\ref{defEtensor}), and $\mu$ is the 
volume form on $\partial \Sigma$, so that $\epsilon_{ab}
=-n_{ab}\wedge \mu$ there.  
The  equality (\ref{eqn:Sbar}) continues to hold at first order in perturbations, which 
can be shown following the same arguments as   given in \cite{Iyer1994a}, hence,
\begin{equation} \label{QWald}
\int_{\partial \Sigma} \delta Q_\zeta = -  \frac{\kappa}{2\pi } \delta S_\text{Wald} \,.
\end{equation}
 The minus sign is opposite the convention in \cite{Iyer1994a} since the unit normal $n^a$
 is outward pointing for the causal diamond.    \\

\paragraph{Generalized volume.} The gravitational part of  $\delta H_\zeta$ is related to the symplectic current $\omega[\delta g, \lie_\zeta g]$   via (\ref{eqn:hamilton}).  The symplectic
form has been computed 
 on an arbitrary background 
for any higher curvature gravitational theory whose Lagrangian is a function 
of the Riemann tensor, but not its covariant derivatives \cite{Bueno:2016ypa}.  Here,
we  take advantage of the maximal symmetry of the background to compute the symplectic
form and Hamiltonian for the causal diamond in any higher order theory, including those 
with  
derivatives of the Riemann tensor.

Recall that the symplectic current $\omega$ is defined in terms of  the symplectic potential $\theta$ through (\ref{eqn:symplectomega}).
For a     Lagrangian that depends on the Riemann tensor and its covariant derivatives, the symplectic potential $\theta^g$ is given in Lemma 3.1 of \cite{Iyer1994a} 
\begin{align}
\theta^g &= 2 E^{bcd}\nabla_d\delta g_{bc}  +S^{ab}\delta g_{ab} \nonumber\\
&+\sum_{i=1}^{m-1} 
T_i^{abcd a_1\ldots a_i} \delta \nabla_{(a_1}\cdots\nabla_{a_i)} R_{abcd} \, ,
\end{align}
where 
$
E^{bcd} = \epsilon_a E^{abcd}
$
and
the tensors $S^{ab}$ and $T_i^{abcda_1\ldots a_i}$ are locally constructed from the metric, 
its curvature, and covariant derivatives of the curvature.  Due to the antisymmetry of 
$E^{bcd}$ in $c$ and $d$, the symplectic current takes the form 
\begin{align}
&\omega^g = 2\delta_1 E^{bcd}\nabla_d\delta_2 g_{bc}-2E^{bcd}\delta_1\Gamma^{e}_{db}
\delta_2 g_{ec} +\delta_1 S^{ab}\delta_2 g_{ab}   \nonumber\\ 
 &+ \sum_{i=1}^{m-1} 
\delta_1T_i^{abcd a_1\ldots a_i} \delta_2 \nabla_{(a_1}\cdots\nabla_{a_i)} R_{abcd} - 
(1\leftrightarrow2) \label{eqn:og}  .
\end{align}
Next we specialize to the geometric setup described in section \ref{subsec:setup}. We may thus employ the fact that we are perturbing around a maximally symmetric background.
This means the background curvature tensor takes the form
\beq\label{msb}
R_{abcd} = \frac{R}{d(d-1)}(g_{ac}g_{bd}-g_{ad}g_{bc})
\eeq
with a constant Ricci scalar $R$, so that $\nabla_e R_{abcd}=0$, and also $\lie_\zeta R_{abcd} \big|_\Sigma = 0$.
Since the tensors $E^{abcd}$, $S^{ab}$, and $T_i^{abcda_1\ldots a_i}$ are 
all constructed from the  metric and  
curvature, they will also have vanishing Lie derivative along
$\zeta^a$ when evaluated on $\Sigma$. 

Replacing $\delta_2 g_{ab}$ in equation (\ref{eqn:og}) with $\lie_\zeta g_{ab}$ and using (\ref{eqn:covlie}),
we obtain
\begin{align} \label{eqn:olzg}
&\omega^g[\delta g, \lie_\zeta g]  \big |_\Sigma =  \nonumber \\
&\qquad\frac2N\left[2 g_{bc}u_d \delta E^{bcd} + E^{bcd}(u_d \delta g_{bc}
- g_{bd}u^e\delta g_{ec}) \right]   .
\end{align}
We would like to write this as a variation of some scalar quantity.  To do so, 
we split off the background value of $E^{abcd}$ by writing
\beq\label{eqn:Fabcd}
F^{abcd} = E^{abcd} - E_0(g^{ac} g^{bd}-g^{ad}g^{bc}) \,.
\eeq
The second term in this expression is the background value, and, due to maximal symmetry, 
the scalar $E_0$ must be a constant determined by the parameters appearing in the Lagrangian.
By definition, $F^{abcd}$ is zero in the background, so any term in (\ref{eqn:olzg}) that depends
on its variation 
may be immediately written as a total variation, since variations of other tensors appearing
in the formula would multiply the background value of $F^{abcd}$, which vanishes.  Hence,
the piece involving $\delta F^{abcd}$ becomes
\beq\label{eqn:dF}
\frac{4}{N} g_{bc}u_d \delta(F^{abcd}\epsilon_a) = \frac{4}{N} 
\delta(F^{abcd} g_{bc}u_d\epsilon_a) \, . 
\eeq
The remaining terms simply involve replacing $E^{abcd}$ in (\ref{eqn:olzg}) with $E_0
(g^{ac}g^{bd}-g^{ad}g^{bc})$.  These terms then take exactly the same form as the 
terms that appear for general relativity, which we know from 
the appendix of \cite{Jacobson:2016aa}
combine to give an overall variation of the volume.  The precise form of this variation when
restricted to $\Sigma$ is 
\beq
-\frac{4(d-2)}{N}\delta \eta \, ,
\eeq
where $\eta$ is the induced volume form on $\Sigma$.  Adding this to (\ref{eqn:dF}) 
produces
\beq \label{eqn:symplform}
\omega[\delta g, \lie_\zeta g]  \big |_\Sigma   = -\frac{4}{N} \delta\left[ \eta(E^{abcd} u_a u_d h_{bc} - E_0) \right] \,,
\eeq
where we used that $\epsilon_a = - u_a \wedge \eta$ on $\Sigma$. This leads us to define a generalized volume functional 
\beq \label{eqn:W}
W = \frac{1}{(d-2)E_0} \int_\Sigma{\eta}(E^{abcd}u_a u_d h_{bc} - E_0) \,,
\eeq
and the variation of this quantity is related to the variation of the gravitational Hamiltonian by 
\beq \label{eqn:dJz}
\delta H_\zeta^g = -4E_0 \kappa k\, \delta W \,,
\eeq
where we have expressed $N$ in terms of $\kappa$ and $k$ using (\ref{eqn:N}).
We have thus succeeded in writing $ \delta H^g_\zeta$ in terms of a 
local geometric functional defined on the surface $\Sigma$. 

It is worth emphasizing that $N$ being constant over the ball was crucial to this derivation, 
since otherwise it could not be pulled out of the 
integral over $\Sigma$ and would define a non-diffeomorphism invariant structure on the surface.
Note that the overall normalization of $W$ is arbitrary, since a different
normalization would simply change the coefficient in front of $\delta W$ in (\ref{eqn:dJz}).  As
one can readily check, the normalization in (\ref{eqn:W}) 
was chosen so that $W$  reduces to the volume in the case of Einstein gravity. In appendix \ref{app:W} we provide   explicit expressions for the generalized volume in $f(R)$ gravity and quadratic gravity.

Finally, combining (\ref{QWald}), (\ref{eqn:dJz}) and (\ref{newvarid}), 
we arrive at the off-shell variational identity in terms of local geometric quantities
\beq \label{eqn:offshelllocalgeo}
\frac{\kappa}{2\pi} \delta S_\text{Wald} -4E_0 \kappa k \delta W + \delta H_\zeta^m = 
\int_\Sigma \delta C_\zeta \, .
\eeq
By imposing the linearized constraints $\delta C_\zeta = 0$, this becomes 
the first law of causal diamond mechanics for higher derivative gravity 
\beq \label{firstlawhigher}
- \delta H_\zeta^m = \frac{\kappa}{2 \pi} \delta S_{\text{Wald}} - 4 E_0 \kappa k  \delta W  \, .
\eeq
This reproduces (\ref{barbecue}) for Einstein gravity with Lagrangian $L = \epsilon R/16\pi G$,
for which $E_0 = 1/32\pi G$.

\subsection{Variation at fixed $W$} \label{sec:fixedflux}
We now show that the first two terms in (\ref{eqn:offshelllocalgeo}) can be written in 
terms of the variation of the Wald entropy at fixed $W$, defined as
\beq \label{eqn:dXbarY}
\delta S_\text{Wald}\big|_{W} = \delta S_\text{Wald} - \frac{\partial S_\text{Wald}}{\partial W}  \delta W \, .
\eeq
Here we must specify what is meant by $\frac{\partial S_\text{Wald}}{\partial W}$.  
We will take this partial
derivative to refer to the changes that occur in both quantities when the size of the ball is 
deformed, but the metric and dynamical fields are held fixed.  Take a vector $v^a$ that is 
everywhere tangent to $\Sigma$ that defines an infinitesimal change in the shape of $\Sigma$.
The first order change this produces in $S_\text{Wald}$ and $W$ can be computed by 
holding $\Sigma$ fixed, but varying the Noether current and Noether charge as $\delta J_\zeta
= \lie_v J_\zeta$ and $\delta Q_\zeta = \lie_v Q_\zeta$.  
Since the background field equations are satisfied  and $\zeta^a$ vanishes on
$\partial \Sigma$, we have there that $\int_{\partial \Sigma}   Q_\zeta = \int_\Sigma J_\zeta^g$,
without reference to the matter part of the Noether current.  
Recall that $\delta W$ is related to the variation of the gravitational Hamiltonian, which can be expressed in terms of $\delta J^g_\zeta$ through (\ref{eqn:hamilton}) and (\ref{eqn:dHz}).
Then using the relations (\ref{eqn:Sbar}) and 
(\ref{eqn:dJz}) and the fact that the Lie
derivative commutes with the exterior derivative, we may compute
\beq
\frac{\partial S_\text{Wald}}{\partial W} = \frac{-\frac{2\pi}{\kappa}\int_{\partial\Sigma} 
\lie_v Q_\zeta}{-\frac{1}{4E_0 \kappa k}\int_\Sigma \lie_v J_\zeta^g}
=  8 \pi E_0 k \,.
\eeq
Combining this result with equations  (\ref{firstlawhigher}) and (\ref{eqn:dXbarY}) we arrive at the off-shell variational identity for higher derivative gravity quoted in the introduction
\beq \label{monkey}
\frac{\kappa}{2\pi} \delta S_\text{Wald}\big|_W +   \delta H_\zeta^m
 = \int_\Sigma \delta C_\zeta \,.
\eeq
Finally, we comment on how JKM ambiguities \cite{Jacobson1994b} 
 affect this identity.  The particular ambiguity we are concerned with
comes from the fact that the symplectic potential $\theta$ in equation (\ref{eqn:dL}) 
is defined only up to 
addition of an exact form $\dd Y[\delta \phi]$ that is linear in the field variations and their 
derivatives. This has the effect of changing the Noether current and Noether charge by
\begin{align}
J_\zeta &\rightarrow J_\zeta + \dd Y[\lie_\zeta \phi] \, ,\\
Q_\zeta &\rightarrow Q_\zeta + Y[\lie_\zeta \phi] \,. \label{eqn:QJKM}
\end{align}
This modifies both the entropy and the generalized volume by  surface terms on $\partial\Sigma$
given by
\begin{align}
   S_{\text{JKM}} &= - \frac{2 \pi}{\kappa} \int_{\partial \Sigma} Y[\lie_\zeta \phi]  \, , \label{eqn:SJKM}\\
  W_\text{JKM} &= - \frac{1}{4 E_0 \kappa k} \int_{\partial\Sigma}  Y[\lie_\zeta \phi] \label{eqn:WJKM} \,.
\end{align}
However, it is clear that this combined change in $J_\zeta$ and $Q_\zeta$ leaves the 
left hand side of (\ref{monkey}) unchanged, since the $Y$-dependent terms cancel out.  
In particular,
\beq
\delta S_\text{Wald}\big|_{W} = \delta(S_\text{Wald}+S_\text{JKM})\big|_{W+W_\text{JKM}} \,,
\eeq
showing that any resolution of the JKM ambiguity gives the same first law, provided that 
   the Wald entropy   and generalized volume   are modified by the terms (\ref{eqn:SJKM}) and 
(\ref{eqn:WJKM}).    This should be expected, because the right hand side of
(\ref{monkey}) depends only on the field equations, which are unaffected by JKM 
ambiguities.


\section{Entanglement Equilibrium}\label{sec:equilibrium}

The original entanglement equilibrium argument for Einstein gravity stated that 
the total variation away from the vacuum of the entanglement of a region at fixed volume is zero.  
This statement is encapsulated in equation (\ref{eqn:dSEEV}), which 
shows both an   area variation due to the change in geometry, and a matter piece
from varying the quantum state.  The area variation at fixed volume can 
equivalently be written  
\beq
\delta A\big|_V = \delta A - \frac{\partial A}{\partial V} \delta V 
\eeq
and the arguments of section \ref{sec:fixedflux} relate this combination to the 
terms appearing in the first law of causal diamond mechanics (\ref{barbecue}).  Since $\delta H_\zeta^m$
in (\ref{barbecue})
is related to $\delta S_\text{mat}$ in (\ref{eqn:dSEEV}) for conformally invariant 
matter, the first law
may be interpreted entirely in terms of entanglement entropy variations.  

This section discusses the extension of the argument to higher derivative theories
of gravity.  Subsection
\ref{sec:subleading} explains how  subleading divergences in the entanglement entropy 
are related to a Wald
entropy, modified by a particular resolution of the JKM ambiguity.  Paralleling the 
Einstein gravity derivation, we seek to relate variations of the subleading 
divergences to the higher derivative
first law of causal diamond mechanics (\ref{firstlawhigher}).  Subsection 
\ref{cons} shows that this can be done as long as the generalized volume $W'$ 
[related to 
$W$ by a boundary JKM term as in (\ref{eqn:Wp})] is held fixed.  Then, using the relation of the 
first law to the off-shell identity (\ref{monkey}), we discuss how the entanglement 
equilibrium condition is equivalent to imposing the linearized constraint equations.

\subsection{Subleading entanglement entropy divergences} \label{sec:subleading}
The structure of  divergences in entanglement entropy is reviewed in \cite{Solodukhin2011a}
and the appendix of \cite{Bousso2016}.  It is well-known that the leading divergence 
depends on the area of the entangling surface.  More surprising, however, is the fact that 
this divergence precisely matches the matter field divergences that renormalize Newton's 
constant.  This ostensible coincidence arises because the two divergences have a 
common origin in the gravitational effective action $I_\text{eff}$, which
includes both gravitational and matter pieces.  Its relation to 
entanglement entropy comes from the replica trick, which defines the entropy as
\cite{Callan1994a,
Calabrese2009} 
\beq \label{eqn:replica}
\see = (n\partial_n - 1)I_\text{eff}(n)\big|_{n=1} \,,
\eeq
where the effective action $I_\text{eff}(n)$ is evaluated on a   manifold with a conical singularity 
at the entangling surface whose excess angle is $2\pi(n-1)$. 

As long as a covariant regulator is used to define the theory, 
the effective action will consist of terms that are local, diffeomorphism invariant
integrals over the manifold, as well as nonlocal contributions. 
All UV matter divergences must appear in the local piece of the effective action, 
and each combines with terms in the classical gravitational part of the action,
renormalizing  the gravitational coupling constants.  Furthermore, each such local
term contributes to the entanglement entropy in (\ref{eqn:replica}) only at the conical 
singularity, giving a local integral over the entangling surface \cite{Fursaev1996,
Larsen1996, Cooperman2013}.  

When the entangling surface is the bifurcation surface
of a stationary horizon, this local integral is simply the Wald entropy 
\cite{Nelson1994, Iyer1995a}. On nonstationary entangling surfaces, 
the computation can be done using the squashed cone techniques of \cite{Fursaev2013a},
which yield terms involving extrinsic curvatures that modify the Wald entropy.  
In holography, the
squashed cone method plays a key role in the proof of the Ryu-Takayanagi
formula \cite{Ryu:2006bv, Lewkowycz2013a}, and its higher curvature generalization
\cite{Dong2013, Camps2013}.  The entropy functionals obtained
in these works seem to also apply outside of holography, giving the 
extrinsic curvature terms in the entanglement entropy for general theories \cite{Fursaev2013a,
Bousso2016}.\footnote{For 
terms involving four or more powers of extrinsic curvature, there are additional
subtleties associated with the so called ``splitting problem'' \cite{Miao2014a, Miao2015a,Camps:2016gfs}. }

The extrinsic curvature modifications to the Wald entropy in fact take the form of a 
JKM Noether charge ambiguity \cite{Jacobson1994b, Sarkar2013, Wall2015}.  
To see this,  note the vector $\zeta^a$  used to define the Noether charge 
vanishes at the entangling surface and its covariant derivative is antisymmetric and 
proportional to the binormal as in equation (\ref{eqn:dz}).  This means it  
acts like a boost on the normal bundle at the entangling surface.  General covariance
requires that any extrinsic curvature contributions can be written as a sum of boost-invariant
products,
\beq
S_\text{JKM} = \int_{\partial\Sigma} \mu \sum_{n\geq 1} B^{(-n)} \cdot C^{(n)}
\eeq
where the superscript $(n)$ denotes the boost weight of that tensor, so that at the surface:
$\lie_\zeta C^{(n)} = n C^{(n)}$.  
It is necessary that the terms consist of two pieces, each of which has nonzero boost weight,
so that they can be written as
\beq
S_\text{JKM} = \int_{\partial\Sigma} \mu \sum_{n\geq1}\frac1n B^{(-n)}\cdot \lie_\zeta C^{(n)} \, .
\eeq
This is of the form of a Noether charge ambiguity from equation (\ref{eqn:QJKM}), 
with\footnote{This formula defines $Y$ at the entangling
surface, and allows for some arbitrariness in defining it off the surface.  It is not clear that $Y$
can always be defined as a covariant functional of the form $Y[\delta\phi, 
\nabla_a\delta\phi,\ldots]$ without reference to additional structures, such as the normal
vectors to the entangling surface.  It would be interesting to understand better if and when 
$Y$ lifts to such a spacetime covariant form off the surface. }
\footnote{We thank
Aron Wall for this explanation of JKM ambiguities.} 
\beq
Y[\delta \phi] = \mu \sum_{n\geq1}\frac1n B^{(-n)} \delta C^{(n)} \,.
\eeq
The upshot of this discussion is that all terms in the entanglement entropy that are local on the
entangling surface, including all divergences, are given by a Wald entropy modified by
specific JKM terms.  
The couplings for the Wald entropy are determined by 
matching to the UV 
completion, or, in the absence of the UV description, these are simply parameters
characterizing the low energy effective theory. 
In induced gravity scenarios, the divergences are determined  by 
the matter content of the theory, and the matching
to gravitational couplings  
has 
been borne
out in explicit examples \cite{Frolov1997, Myers2013, Pourhasan2014}.

\subsection{Equilibrium condition as gravitational constraints}\label{cons}

We can now relate the variational identity (\ref{monkey}) to entanglement entropy.  
The reduced density matrix for the ball in vacuum takes the form
\beq
\rho_\Sigma = e^{- H_\text{mod}}/Z \,,
\eeq
where $H_\text{mod}$ is the modular Hamiltonian
and $Z$ is the partition function, ensuring
that $\rho_\Sigma$ is normalized.  
Since the matter is conformally invariant, the 
modular Hamiltonian takes a simple form in terms of the matter Hamiltonian $H_\zeta^m$
defined in (\ref{potati}) \cite{Hislop1982, Casini2011}
\beq \label{eqn:Hmod}
H_\text{mod} 
= \frac{2\pi}\kappa H_\zeta^m \,. 
\eeq 
Next we apply the first law of 
entanglement entropy \cite{Blanco2013a, Bhattacharya2012}, 
which states that the first order perturbation to the
entanglement entropy is given by the change in modular Hamiltonian expectation value
\beq
\delta S_\text{EE} =  \delta\vev{H_\text{mod}} \, .
\eeq
Note that this equation holds for a fixed geometry and entangling surface, and hence
coincides with what was referred to as $\delta S_\text{mat}$ in section \ref{sec:intro}.   When varying the 
geometry, the divergent part of the entanglement entropy changes due to a 
change in the Wald entropy and JKM terms of the entangling surface.  
The total variation of the entanglement entropy is therefore
\beq \label{eqn:dSEEtot}
\delta S_\text{EE}=  \delta (S_\text{Wald}+S_\text{JKM})  + \delta\vev{H_\text{mod}} \, .
\eeq
At this point, we must give a prescription for defining the surface $\Sigma$ in the perturbed
geometry.  Motivated by the first law of causal diamond mechanics, we  require that $\Sigma$ 
has the same generalized volume $W'$ as in vacuum, where $W'$ differs from 
the  quantity  $W$ by a JKM term, as in equation (\ref{eqn:Wp}). This provides
a diffeomorphism-invariant criterion for defining the size of the ball.  It does
not fully fix all properties of the surface, but it is enough to derive the equilibrium
condition for the entropy.  As argued in section \ref{sec:fixedflux}, the first term in 
equation (\ref{eqn:dSEEtot}) can be written instead as $\delta S_\text{Wald}\big|_W$ 
when the variation is taken holding $W'$ fixed. 
Thus, from equations (\ref{monkey}), 
(\ref{eqn:Hmod}) and (\ref{eqn:dSEEtot}), we arrive at our main result, the equilibrium condition
\beq \label{eqn:dSEEW}
\frac{\kappa}{2\pi}\delta S_{\text{EE}} \big|_{W'} = \int_\Sigma \delta C_\zeta \,,
\eeq
valid for minimally coupled, conformally invariant matter fields.

The linearized constraint equations $\delta C_\zeta=0$ may therefore be interpreted as 
an equilibrium condition on entanglement entropy for the vacuum.  
Since all first variations of the entropy vanish when the linearized gravitational constraints 
are 
satisfied, the vacuum is an extremum of entropy for 
regions with fixed generalized volume $W'$, which is necessary for it
to be an equilibrium 
state. 
Alternatively, postulating that entanglement entropy is maximal in vacuum
for all balls and in all frames would allow one to conclude that the linearized higher
derivative equations hold everywhere.

\section{Field equations from the equilibrium condition} \label{sec:equations}

The entanglement equilibrium hypothesis provides a clear connection between the 
linearized gravitational constraints and the maximality of entanglement entropy at 
fixed $W'$ in the vacuum for conformally invariant matter.  In this section, 
we will
consider whether information about the fully nonlinear 
field equations can be gleaned from the equilibrium condition. Following the approach
taken in \cite{Jacobson:2016aa}, we employ
a limit where the ball is taken to be much smaller than all relevant scales in the problem, but
much larger than the cutoff scale of the effective field theory, which is  set by the 
gravitational coupling constants.
By expressing the linearized equations in Riemann normal
coordinates, one can infer that the full \emph{nonlinear} field equations hold in the 
case of Einstein gravity.
As we discuss here, such a conclusion can \emph{not} be reached for higher curvature
theories.  The main issue is that higher order terms in the RNC expansion are needed to 
capture the effect of higher curvature terms in the field equations, but these contribute
at the same order as nonlinear corrections to the linearized equations.

We begin by reviewing the argument for Einstein gravity.
Near any given point, the metric looks locally flat, and has an expansion in 
terms of Riemann normal coordinates that takes the form
\beq
g_{ab}(x) = \eta_{ab}-\frac13x^c x^d R_{acbd}(0) +\mathcal{O}(x^3)\, ,
\eeq
where   $(0)$ means evaluation at the center of the ball. 
At distances small compared to the radius of curvature, the second term in this expression
is a small perturbation to the flat space metric $\eta_{ab}$.  Hence, we may apply 
the off-shell identity (\ref{eqn:dSEEW}), using the first order variation
\beq \label{eqn:dgRNC}
\delta g_{ab} = -\frac13 x^c x^d R_{acbd} (0) \,,
\eeq
and conclude that the linearized constraint $\delta C_\zeta$ 
holds for this metric perturbation.  When 
restricted to the surface $\Sigma$, 
this constraint in Einstein gravity is \cite{Seifert2007b}
\beq
C_\zeta\big|_\Sigma=-u^a \zeta^b\left(\frac1{8\pi G} G_{ab}-T_{ab}\right) \eta \,.
\eeq 
Since the background constraint is assumed to hold, the perturbed constraint is 
\beq
\delta C_\zeta\big|_\Sigma = -u^a \zeta^b\left(\frac{1}{8\pi G} \delta G_{ab}- \delta T_{ab}\right)\eta\, ,
\eeq
but in Riemann normal coordinates, we have that the linearized perturbation to the curvature is
just $\delta G_{ab} = G_{ab}(0)$, up to terms suppressed by the ball radius.  
Assuming that the ball is small enough so that the stress tensor
may be taken constant over the ball, one concludes that the vanishing constraint implies the 
nonlinear field equation  at the center of the ball\footnote{In this equation, 
$\delta T_{ab}$ should be thought of as a quantum expectation value of the 
stress tensor.  Presumably, for sub-Planckian energy densities and in the small ball limit, this first order
variation approximates the true energy density.  However, 
there exist states for which the change in stress-energy is zero at first order in perturbations
away from the vacuum, most notable for coherent states \cite{Varadarajan2016a}.  
Analyzing how these states can be incorporated into the entanglement equilibrium 
story deserves further attention. }
\beq
u^a \zeta^b(G_{ab}(0) - 8\pi G \delta T_{ab}) = 0\,.
\eeq
The procedure outlined above applies at all points and all frames, allowing us to obtain the full tensorial Einstein equation.  

Since we have only been dealing with the linearized constraint, one could question
whether it gives a good approximation to the field equations at all points within 
the small ball.  This
requires estimating the size of the nonlinear corrections to this field equation.   When 
integrated over the ball, the corrections to the curvature in RNC 
are of order $\ell^2/L^2$, where 
$\ell$ is the radius of the ball and $L$ is the radius of curvature.  Since we took the ball
size to be much smaller than the radius of curvature, these terms are already suppressed
relative to the linear order terms in the field equation.  

The situation in higher derivative theories of gravity is much different.
 It is no longer the case that the linearized equations evaluated in RNC imply the full nonlinear 
field equations   in a small ball.
To see this,  consider an $L[g_{ab},R_{bcde}]$  
higher curvature theory.\footnote{Note that an analogous argument should hold for general higher derivative theories, which also involve covariant derivatives of the Riemann tensor.} The equations of motion read
\begin{equation}\label{eomhigh}
- \frac{1}{2} g^{ab} \mathcal L  + E^{aecd} \tensor{R}{^b_{ecd}} - 2  \nabla_c \nabla_d E^{acdb}
= \frac{1}{2} T^{ab} \,.
\end{equation}
In appendix \ref{app:FLDMRNC} we show that linearizing these equations
around a Minkowski background leads to
\begin{align}\label{lineomhigh}
 \frac{\delta G^{ab}}{16\pi G} - 2  \partial_c \partial_d \delta E^{a c d b}_{\text{higher}} = \frac{1}{2} \delta T^{ab}\, ,
\end{align}
where we split $E^{abcd}=E^{abcd}_{\text{Ein}}+E^{abcd}_{\text{higher}}$ into its Einstein piece, which gives rise to the Einstein tensor, and a piece coming from higher derivative terms.
As noted before, the variation of the Einstein tensor evaluated in RNC gives the nonlinear Einstein tensor, up to corrections that are suppressed by the ratio of the ball size to the radius of curvature. 
However, in a higher curvature theory of gravity, the equations of motion 
\eqref{eomhigh} contain terms that are nonlinear in the curvature.
Linearization around a MSS background of these terms would 
produce, schematically, $\delta ( R^n ) = n \bar{R}^{n-1} \delta R$, where $\bar{R}$ denotes evaluation in the MSS background.
In Minkowski space, all such terms would vanish.
This is not true in a general MSS, but evaluating the curvature tensors in the 
background still leads to a significant loss of information about the tensor structure 
of the equation.  
We conclude that the linearized equations cannot reproduce the full nonlinear 
field equations for higher curvature gravity, and it is only the linearity of the Einstein equation
in the curvature that allows the nonlinear equations to be obtained for general relativity.

When linearizing around flat space, the higher curvature corrections to the Einstein equation  are entirely captured by the  second term in \eqref{lineomhigh}, which features  
four derivatives acting on the metric, since $E_\text{higher}^{abcd}$ is constructed from 
curvatures that already contain two derivatives of the metric.
Therefore, one is insensitive to higher curvature corrections unless at least $\mathcal{O}(x^4)$ corrections \cite{Brewin2009} are added to the Riemann normal coordinates expansion \eqref{eqn:dgRNC}
\beq\label{eqn:dg4}
\delta g^{(2)}_{ab} = x^c x^d x^e x^f\bigg(\frac2{45}R\indices{_a _c_d^g}R\indices{_b_e_f_g} -\frac1{20}\nabla_c\nabla_d R_{aebf}\bigg) \,.
\eeq
Being quadratic in the Riemann tensor, this term contributes at the same order as 
the nonlinear corrections to the linearized field equations. Hence, linearization based on 
the RNC expansion up to $x^4$ terms is not fully self-consistent.
This affirms the claim that for higher curvature theories, the nonlinear equations at a point cannot be derived by only imposing the linearized equations.


 \section{Discussion} \label{sec:conclusion}
 Maximal entanglement of the vacuum state was proposed in \cite{Jacobson:2016aa} as a new 
 principle in quantum gravity.  It hinges on the assumption that divergences in the 
 entanglement entropy are cut off  at short distances, so it
 is ultimately a statement about the UV complete quantum gravity theory.  However, the 
 principle can be phrased in terms of the generalized entropy, which is intrinsically UV 
 finite and well-defined within the low energy effective theory.  
 Therefore, if true, maximal vacuum entanglement provides a 
 low energy constraint on any putative
 UV completion of a gravitational effective theory.  
 
 Higher curvature terms arise generically in any such effective  field theory.  Thus, it
 is important to understand how the entanglement equilibrium argument is modified by 
 them.  
As explained in section \ref{sec:firstlaw}, the precise characterization of the entanglement equilibrium hypothesis  
relies on a classical variational identity for causal diamonds in maximally symmetric spacetimes.
This identity leads to equation (\ref{monkey}), which relates variations of the Wald entropy and 
matter energy density of the ball to the linearized constraints. The variations
are taken holding fixed a new geometric
functional $W$, defined in (\ref{eqn:W}), which we call the ``generalized volume.''

We connected this identity to entanglement equilibrium in section \ref{sec:equilibrium},
invoking the   fact that subleading entanglement entropy divergences are given by a Wald
entropy, modified by specific JKM terms, which also modify $W$ by the boundary
term (\ref{eqn:WJKM}).  
  With the additional assumption that matter is conformally invariant, we arrived at our main result \req{eqn:dSEEW}, showing that the equilibrium condition $\delta S_\text{EE}\big|_{W'}=0$ 
  applied to small balls is
  equivalent to imposing the linearized constraints $\delta C_\zeta = 0$.

In section \ref{sec:equations}, we reviewed the argument that
in the special case of Einstein gravity, 
 imposing the linearized equations within small enough balls is equivalent to 
requiring that the fully nonlinear equations hold within the ball \cite{Jacobson:2016aa}. Thus by considering spheres
centered at each point and in all Lorentz frames, one could 
conclude that the full Einstein equations hold everywhere.\footnote{There
is a subtlety associated with whether the solutions within each small
ball can be consistently glued together to give a solution over all of spacetime.  
One must solve for the gauge transformation relating the Riemann normal 
coordinates at different nearby points, and errors in the linearized approximation
could accumulate as one moves from point to point.  The question of whether the ball
size can be made small enough so that the total accumulated error goes to zero
deserves further attention.}  Such an argument cannot be
made for a theory that involves higher curvature terms.  One finds that higher order terms
in the RNC expansion are needed to detect the higher 
curvature pieces of the field equations, but these terms enter at the same order as 
the nonlinear corrections to the linearized equations. This signals 
a breakdown of the perturbative
expansion unless the curvature is small.   

The fact that we obtain only linearized equations for the higher curvature theory 
is consistent with the effective field theory standpoint.  One could take 
the viewpoint that higher curvature corrections are 
 suppressed by powers of a UV scale, and the effective field theory is valid 
only when the curvature is small compared to this scale.  This suppression would suggest
that the linearized equations largely capture the effects of the higher curvature corrections
in the regime where effective field theory is reliable.

\subsection{Comparison to other ``geometry from entanglement'' approaches}
Several proposals have been put forward to understand gravitational dynamics
in terms of thermodynamics and entanglement.  Here we will compare the 
entanglement equilibrium program considered in this paper 
to two other approaches: the equation of state for
local causal horizons, and gravitational dynamics from holographic entanglement
entropy  (see \cite{Carroll2016a} for a related discussion).

\subsubsection{Causal horizon equation of state}
By assigning an entropy proportional to the area of local causal horizons, Jacobson
showed that the Einstein equation arises as an equation of state  
\cite{Jacobson1995a}.  This approach employs a physical process first law for the local 
causal horizon, defining a heat $\delta Q$ as the flux of local boost energy across the horizon.  
By assigning an entropy $S$ to the horizon proportional to its area, one finds that the 
Clausius relation $\delta Q = T\delta S$ applied to all such horizons is equivalent to the Einstein
equation.

The entanglement equilibrium approach differs in that it employs an equilibrium state first 
law [equation (\ref{firstlawhigher})], instead of  a physical process one \cite{Wald1994}.  
It therefore represents 
a different perspective that focuses on the steady-state behavior, as opposed to dynamics
involved with evolution along the causal horizon.
It is consistent therefore that we obtain constraint
equations in the entanglement equilibrium setup, since one would not expect evolution
equations to arise as an equilibrium condition.\footnote{We thank Ted Jacobson
for clarifying this point.}   That we can infer dynamical equations
from the constraints is related to the fact that the dynamics of diffeomorphism-invariant 
theories is entirely determined by the constraints evaluated in all possible Lorentz frames.  

Another difference comes from the focus on spacelike balls as opposed to local causal 
horizons.  Dealing with a compact spatial region has the advantage of providing an IR 
finite entanglement entropy, whereas the entanglement associated with local causal
horizons can depend on fields far away from the point of interest.  This allows us to give 
a clear physical interpretation for the surface entropy functional as entanglement entropy, 
whereas such an interpretation is less precise in the equation of state approaches.  

Finally, we note that both approaches attempt to obtain fully nonlinear equations by
considering ultralocal regions of spacetime.  In both cases the derivation of the field equations
for Einstein gravity is fairly robust, however higher curvature corrections present some problems.
Attempts have been made in the local causal horizon approach
that involve modifying
the entropy density functional for the horizon
\cite{Eling2006c, Elizalde2008, Chirco2011,
 Brustein2009, Parikh2016, Dey2016, Padmanabhan2009, Padmanabhan2009a,
 Guedens2011}, but they meet certain challenges.  
 These include a need for a physical interpretation of the chosen entropy density functional, 
 and dependence of the entropy on arbitrary features of the local
Killing vector in the vicinity of the horizon
 \cite{Guedens2011, Jacobson2012b}.  
While the entanglement equilibrium argument avoids these problems, it fails to get beyond
linearized higher curvature equations, even after considering the small ball limit.  The 
nonlinear equations in this case appear to involve information beyond first order perturbations,
and hence may not be accessible based purely on an equilibrium argument.

\subsubsection{ Holographic entanglement entropy }
A different approach  comes from holography and the 
Ryu-Takayanagi formula \cite{Ryu:2006bv}.  By demanding that areas of
minimal surfaces
in the bulk match the entanglement entropies of spherical regions in the boundary CFT, one
can show that the linearized gravitational equations must hold \cite{Lashkari2013, Faulkner2013,
Swingle2014a}.  The argument employs an equilibrium state first law for the bulk geometry,
utilizing the Killing symmetry associated with Rindler wedges in the bulk.

The holographic approach is quite similar to the entanglement equilibrium argument since
both use equilibrium state first laws.  One difference is that the holographic argument must 
utilize minimal surfaces in the bulk, which extend all the way to the boundary of AdS.  This 
precludes using a small ball limit as can be done with the entanglement 
equilibrium derivation, and is the underlying reason that entanglement equilibrium can derive fully
nonlinear field equations in the case of Einstein gravity, 
whereas the holographic approach has thus far only obtained linearized equations.  Some progress has been made to go beyond 
linear order in the holographic approach by considering higher order perturbations
in the bulk \cite{Faulkner2014, Lashkari2015, Beach2016}.  Higher order 
perturbations will prove useful in the entanglement equilibrium program as well,
and has the potential to extend the higher curvature derivation to fully nonlinear equations.
Due to the similarity between the holographic and entanglement equlibrium approaches, 
progress in one will complement and inform  the other.

\subsection{Thermodynamic interpretation of the first law of causal diamond mechanics}
Apart from the entanglement equilibrium interpretation, the first law of causal diamond mechanics could also directly be interpreted  as a thermodynamic relation.   Note that the identity (\ref{barbecue}) for Einstein gravity bears a  striking resemblance  to the fundamental relation in thermodynamics
\beq \label{firstlawthermo}
dU = T dS - p dV,
\eeq
where $U(S,V)$ is the internal energy, which is a function of the   entropy $S$  and volume $V$.  
The  first law  (\ref{barbecue})  turns into the thermodynamic   relation (\ref{firstlawthermo}),   if one makes the following identifications for the temperature $T$ and pressure $p$
\beq \label{temppress}
T = \frac{\kappa \hbar}{2 \pi k_B c} \, , \quad \quad p = \frac{c^2 \kappa k}{8 \pi G} \, .
\eeq
Here we have restored fundamental constants, so that the quantities on the RHS have  the standard  units of temperature and pressure.
The expression for the temperature is the well-known Unruh    temperature  \cite{Unruh1976}. The formula for the pressure lacks a microscopic understanding at the moment, although we emphasize the expression follows from consistency of the first law.

The thermodynamic interpretation   motivates the name ``first law'' assigned to (\ref{barbecue}), and  arguably it   justifies the terminology ``generalized volume'' used for $W$   in this paper, since it enters into the first law for higher curvature gravity (\ref{firstlawhigher}) in the place of the volume.
The only difference with the fundamental relation in thermodynamics is the minus sign in front of the energy variation. This different sign  also enters into the first law for   de Sitter horizons \cite{Gibbons1977}. In the latter case the sign appears because   empty de Sitter spacetime has maximal entropy, and adding matter only decreases the horizon entropy. Causal diamonds are rather similar in that respect.

\subsection{Generalized volume and holographic complexity} 
The emergence of  a generalized notion of volume in this analysis is interesting in
its own right.  We showed that when perturbing around a maximally symmetric background, the 
variation of the generalized volume is proportional to the variation of the gravitational part of the Hamiltonian.  The fact that the Hamiltonian could be 
written in terms of a local, geometric functional of the surface was a nontrivial consequence of 
the background geometry being maximally symmetric and $\zeta^a$ being a conformal
Killing vector whose conformal factor vanishes on $\Sigma$.  The local geometric nature
of $W$ makes it a useful, diffeomorphism invariant quantity with which to characterize the 
region under consideration, and thus should be a good state function in the thermodynamic 
description of an ensemble of quantum geometry microstates.  One might hope that such
a microscopic description would also justify the fixed-$W'$ constraint 
in the entanglement equilibrium derivation, which was only  motivated macroscopically 
by the first law of causal diamond mechanics.

Volume has  recently been identified as an important quantity in holography, where it is
conjectured to be related to complexity \cite{Susskind2014,Brown2015a}, or 
fidelity susceptibility \cite{Miyaji2015}.  The complexity$=$volume conjecture states that the complexity of some boundary state on a
time slice $\Omega$ is proportional to the  volume of the extremal codimension-one bulk hypersurface $\mathcal{B}$ which meets the asymptotic boundary on the corresponding time slice.\footnote{A similar expression has also been proposed for the complexity of subregions of the boundary time slice. In that case, $\mathcal{B}$ is the bulk hypersurface bounded by the corresponding subregion on the asymptotic boundary and
the Ryu-Takayanagi surface \cite{Ryu:2006bv} in the bulk \cite{Alishahiha2015, Ben-Ami2016},
or, more generally, the Hubeny-Rangamani-Takayanagi surface \cite{Hubeny:2007xt} if the spacetime is time-dependent \cite{Carmi2016}.
} 

While volume is the natural functional to consider for Einstein gravity, \cite{Alishahiha2015} noted that
this should be generalized for higher curvature theories.  The functional 
proposed in that work resembles our generalized volume $W$, but suffers from 
an arbitrary dependence on the choice of foliation of the codimension-one hypersurface
on which it is evaluated.  We therefore suggest that $W$, as defined in \req{eqn:W}, may provide a suitable generalization
of volume in the context of higher curvature holographic complexity.

Observe however that our derivation of $W$ using the Iyer-Wald formalism was carried out in the particular case of spherical regions whose causal diamond is preserved by a conformal Killing vector. On more general grounds, one could speculate that
the holographic complexity functional in higher derivative gravities should involve contractions of $E^{abcd}$ with the geometric quantities characterizing $\mathcal{B}$, namely the  induced  metric $h_{ab}$ and the normal vector $u^a$.
The most general functional involving at most one factor of $E^{abcd}$ can be written as
\begin{equation}\label{wwe}
\mathcal{W}(\mathcal{B})=\int_{\mathcal{B}} \eta \left(\alpha E^{abcd}u_a h_{bc}u_d+ \beta E^{abcd} h_{ad}h_{bc}+\gamma\right)\, ,
\end{equation}
for some constants $\alpha$, $\beta$ and $\gamma$ which should be such that $\mathcal{W}(\mathcal{B})=V(\mathcal{B})$ for Einstein gravity. It would be interesting to explore the validity of this proposal in particular holographic setups, e.g., along the lines of \cite{Carmi2016}.

\subsection{Future work}
We conclude by laying out future directions for the entanglement equilibrium program.

\subsubsection{Higher order perturbations}
In this work we restricted attention only to first order perturbations of the entanglement
entropy and the geometry.  Working to higher order in perturbation theory could yield several
interesting results.  One such possibility would be proving that the vacuum entanglement 
entropy is maximal, as opposed to merely extremal.  The second order change in entanglement
entropy is no longer just the change in modular Hamiltonian expectation value.  The 
difference is given by the relative entropy, so a proof of maximality will likely invoke
the positivity of relative entropy.  On the geometrical side, a second order variational
identity would need to be derived, along the lines of \cite{Hollands2013}.  
One would expect that graviton contributions would appear at this order, and it 
would be interesting to examine how they play into the entanglement equilibrium story.  
Also, by considering small balls and using the higher order terms in the Riemann normal
coordinate expansion
(\ref{eqn:dg4}), in addition to   higher order perturbations,  it is possible that one could derive the fully nonlinear
field equations of any higher curvature theory.  Finally, coherent states pose a puzzle
for the entanglement equilibrium hypothesis, since they change the energy within
the ball without changing the entanglement \cite{Varadarajan2016a}.  However, 
their effect on the energy density only appears at second order in perturbations, so 
carrying the entanglement equilibrium argument to higher order could shed light on 
this puzzle.

\subsubsection{Nonconformal matter} \label{sec:nonconformal}
The arguments deriving the entanglement equilibrium condition in section \ref{cons}
were restricted to matter that is conformally invariant.  For nonconformal matter, there are 
corrections to the modular Hamiltonian that spoil the relation between $\delta S_\text{mat}$
and the matter Hamiltonian $H_\zeta^m$.  Nevertheless, in the small ball limit these corrections
take on a simple form, and one possible solution for extending the entanglement equilibrium 
argument  introduces a local cosmological constant to absorb the effects of
the modular Hamiltonian corrections \cite{Jacobson:2016aa, Casini2016, Speranza2016}.
Allowing variations of the local cosmological constant would result in a modified first 
law \cite{Manus}, and may have connections to the black hole chemistry program 
\cite{Kastor2009, Kubiznak2016}.
It is also possible that some other resolution exists to this apparent conflict, perhaps involving 
the RG properties of the matter field theory when taking the small ball limit.

\subsubsection{Nonminimal couplings and gauge fields} \label{sec:nonmin}
 We restricted attention to minimally coupled matter throughout this work.  Allowing for 
 nonminimal coupling can lead to new, state-dependent divergences in the entanglement
 entropy \cite{Marolf2016}.  As before, these divergences  will be localized on 
 the entangling surface, taking the form of a Wald entropy.  It therefore
  seems plausible that an entanglement equilibrium argument will go through in this
 case, reproducing the field equations involving the nonminimally coupled field.  
 Note the state-dependent divergences could lead to variations of the couplings in the 
 higher curvature theory, which may connect to the entanglement chemistry program,
 which considers Iyer-Wald first laws involving variations of the couplings \cite{Caceres2016}.

Gauge fields introduce additional subtleties related
 to the existence of edge modes \cite{Donnelly2012, Donnelly2015, Donnelly2016}, and since
 these affect the renormalization of the gravitational couplings, they require special attention.  
 Gravitons are even more problematic due to difficulties in defining the entangling surface in
 a diffeomorphism-invariant manner and in finding a covariant regulator \cite{Fursaev1997,
 Cooperman2013, Solodukhin2015, Bousso2016}.  It would be 
 interesting to analyze how to handle these issues in the entanglement equilibrium
 argument.

\subsubsection{ Nonspherical subregions}
 The entanglement equilibrium condition was shown to hold for spherical subregions and 
 conformally invariant matter.  One question that arises is whether an analogous 
 equilibrium statement holds for linear perturbations to the vacuum in an arbitrarily
 shaped region.  Nonspherical regions present a challenge because
there is no longer a simple relation between the modular Hamiltonian and the matter stress
tensor.  Furthermore, nonspherical regions do not admit a conformal Killing vector 
which preserves its causal development.  Since many properties of the conformal
Killing vector were used when deriving the generalized volume $W$, 
it may 
need to be modified to apply to nonspherical regions and their perturbations.

Adapting the entanglement equilibrium arguments to nonspherical regions may involve shifting 
the focus to evolution under the modular flow, as opposed to a geometrical evolution generated
by a vector field.  Modular flows are complicated in general, but one may be able to use general
properties of the flow to determine whether the Einstein equations still imply maximality
of the vacuum entanglement for the region.  Understanding the modular flow may also
shed light on the behavior of the entanglement entropy for nonconformal matter, and whether
some version of the entanglement equilibrium hypothesis continues to hold.

\subsubsection{ Physical process}
As emphasized above, the first law of causal diamond mechanics is an equilibrium state construction
since it compares the entropy of  $\partial \Sigma$ on two infinitesimally related geometries
\cite{Wald1994}.  One could 
ask whether there exists a physical process version of this story, which
deals with entropy changes and energy fluxes as you evolve along the null boundary of the 
causal diamond.  For this, the notion of quantum expansion for the null surface
introduced in \cite{Bousso2016} would be a useful concept, which is defined by the derivative
of the generalized entropy along the generators of the surface.  One possible subtlety in 
formulating a physical process first law for the causal diamond is that the (classical) 
expansion 
of the null boundary is nonvanishing, so it would appear that this setup does not
correspond to a 
dynamical equilibrium configuration.  Nevertheless, it may be possible to gain useful information
about the dynamics of semiclassical gravity by considering these nonequilibrium
physical processes. 
 An alternative that avoids this issue 
is to focus on quantum extremal surfaces \cite{Engelhardt2015} whose quantum expansion 
vanishes, and therefore may lend themselves to an equilibrium physical process
first law.

\begin{acknowledgments}
We would like to thank Joan Camps, 
Ted Jacobson, Arif Mohd, Rob Myers, Erik Verlinde and  Aron Wall    for helpful
discussions, Fernando Rejon-Barrera for an early collaboration on this project, 
and Ted Jacobson for comments on a draft of this work.   
AJS is grateful to the Maryland Center for Fundamental Physics and Aron Wall
for organizing the ``Minicourse on Spacetime Thermodynamics.''  VSM, AJS and MRV 
thank the organizers of the ``Amsterdam String Workshop,'' hosted by the Delta Institute
for Theoretical Physics, and PB and AJS are grateful to the organizers of the ``It from 
Qubit Summer School'' held at the Perimeter Institute for Theoretical Physics. Research at Perimeter Institute is supported
by the Government of Canada through Industry Canada and by the Province of Ontario
through the Ministry of Research and Innovation.
The work of PB is supported by a postdoctoral fellowship from the Fund for Scientific Research - Flanders (FWO). PB also acknowledges support from the Delta ITP Visitors Programme. VSM is supported by a PhD fellowship from the FWO and
  by the ERC grant  616732-HoloQosmos. AJS is supported by the National Science Foundation  under 
grant No.\ PHY-1407744.    
  MRV acknowledges support from  the ERC Advanced Grant 268088-EMERGRAV, the Spinoza Grant of the Dutch Science Organisation (NWO), and    the NWO Gravitation Program for the Delta Institute for Theoretical Physics.
\end{acknowledgments}
 
 \appendix

\section{Conformal Killing vector in flat space}  \label{appkill}
Here we make explicit the geometric quantities introduced in section \ref{subsec:setup} in the case of a Minkowski background, whose metric we write in spherical coordinates, i.e., $ds^2 = - dt^2 + dr^2 + r^2 d \Omega_{d-2}^2$. Let $\Sigma$ be a spatial ball of radius $\ell$ in the time slice $t=0$ and with center at $r=0$. The conformal Killing vector which preserves the causal diamond  of $\Sigma$ is given by \cite{Jacobson:2016aa} 
\begin{equation}\label{zz}
\zeta=\left( \frac{\ell^2-r^2-t^2}{\ell^2}\right) \partial_t-\frac{2 r t}{\ell^2} \partial_r \, ,
\end{equation}
where we have chosen the normalization in a way such that $\zeta^2=-1$ at the center of the ball, which then gives the usual notion of energy for $H_\zeta^m$ (i.e. the correct units).
It is straightforward to check that 
$
\zeta(t=\pm \ell,r=0)=\zeta(t=0,r=\ell)=0\, ,
$
 i.e., the tips of the causal diamond and the maximal sphere $\partial \Sigma$ at its waist are fixed points of $\zeta$, as expected. Similarly, $\zeta$ is null on the boundary of the diamond. In particular, 
 $
 \zeta(t=\ell \pm r)=\mp 2 r(\ell \pm r)/\ell^2 \cdot (\partial_t \pm \partial_r)\, .
$
The vectors $u$ and $n$ (respectively normal to $\Sigma$ and to both $\Sigma$ and $\partial \Sigma$) read
$u=\partial_t$, $n=\partial_r$,
so that the binormal to $\partial \Sigma$ is given by 
$
n_{ab}=2\nabla_{[a} r \nabla_{b]} t \, .
$
It is also easy to check that $\lie_{\zeta}g_{ab}=2\alpha g_{ab}$ holds, where
$
\alpha\equiv \nabla_{a}\zeta^a/d=-2t/\ell^2 \,.
$
Hence, we immediately see that $\alpha=0$ on $\Sigma$, which implies that the gradient of $\alpha$ is proportional to the unit normal $u_a=-\nabla_a t$. Indeed, one finds $ \nabla_a \alpha=-2\nabla_a t/\ell^2 $, 
so in this case $N\equiv \lVert \nabla_a \alpha \rVert^{-1}=\ell^2/2$. It is also easy to show that $(\nabla_a \zeta_b)|_{\partial \Sigma}=\kappa n_{ab}$ holds, where the surface gravity reads $\kappa=2/\ell$.

As shown in \cite{Jacobson1993}, given some metric $g_{ab}$ with a conformal Killing field $\zeta^a$, it is possible to construct other metrics $\bar{g}_{ab}$ conformally related to it, for which $\zeta^a$ is a true Killing field. More explicitly, if 
$\lie_{\zeta}g_{ab}=2\alpha g_{ab}$, then $\lie_{\zeta}\bar{g}_{ab}=0$ 
as long as $g_{ab}$ and $\bar{g}_{ab}$ are related through $\bar{g}_{ab}=\Phi\, g_{ab}$, where $\Phi$ satisfies
\begin{equation}
 \lie_{\zeta} \Phi+2\alpha \Phi=0\, .
\end{equation}
For the vector \req{zz}, this equation has the general solution 
\begin{equation}\label{pii}
\Phi(r,t)=\frac{\psi(s)}{r^2}\, \quad  \text{where} \quad s\equiv \frac{\ell^2+r^2-t^2}{r} \, .
\end{equation}
Here, $\psi(s)$ can be any function. Hence, $\zeta$ in \req{zz} is a true Killing vector for all metrics conformally related to   Minkowski's  with a conformal factor given by \req{pii}. For example, setting $\psi(s)=L^2$, for some constant $L^2$, one obtains the metric of AdS$_2\times S_{d-2}$ with equal radii, namely: 
$ds^2=L^2/r^2 (-dt^2+dr^2)+L^2 d\Omega_{d-2}^2
$. Another simple case corresponds to $\psi(s)=L^2((s^2/(4L^2)-1)^{-1}$.  Through the change of variables \cite{Casini2011}: $t=L \sinh(\tau/L)/(\cosh u +\cosh (\tau/L))$, $r=L \sinh u \, /(\cosh u +\cosh (\tau/L))$, this choice leads to the $\mathbb{R}\times H^{d-1}$ metric (where $H^{d-1}$ is the hyperbolic plane): $ds^2= -d\tau^2+L^2(du^2+\sinh^2 u \, d\Omega^2_{d-2})$. 


\section{Generalized volume in higher order gravity} \label{app:W}

\noindent The generalized volume $W$ is defined in \eqref{eqn:W}.
We restate the expression here
\beq
W= \frac{1}{(d-2)E_0} \int_\Sigma \eta    \left (  E^{abcd} u_a u_d h_{bc}  - E_0  \right) \, ,
\eeq
where $E_0$ is   a theory-dependent constant defined by the tensor $E^{abcd}$
in a maximally symmetric solution to the field equations through
$E^{abcd}\overset{\text{MSS}}{=}E_0(g^{ac}g^{bd}-g^{ad}g^{bc})$. Moreover, $E^{abcd}$     is  the variation of the Lagrangian scalar $\mathcal L$ with respect to the Riemann tensor $R_{abcd}$ if we were to treat it as an independent field \cite{Iyer1994a},
\begin{align} \label{defEtensor}
E^{abcd} &= \frac{\partial   \mathcal L}{\partial R_{abcd}}  - \nabla_{a_1} \frac{\partial \mathcal L}{\partial \nabla_{a_1} R_{abcd}} + \dots   \\
&+ (-1)^m \nabla_{(a_1} \cdots \nabla_{a_m )} \frac{\partial \mathcal L}{\partial  \nabla_{(a_1} \cdots \nabla_{a_m )} R_{abcd}} \nonumber  \, ,
\end{align}
where $\mathcal L$ is then  defined  through $L = \epsilon \mathcal L$. In this section we provide explicit expressions for $W$ in $f(R)$ gravity, quadratic gravity and Gauss-Bonnet gravity. Observe that throughout this section we use the bar on $\bar{R}$ to denote evaluation on a MSS. Imposing a MSS to solve the field equations of a given higher derivative theory gives rise to a constraint between the theory couplings and the background curvature $\bar{R}$. This reads \cite{Bueno:2016ypa}
\begin{equation}\label{emb}
E_0=\frac{d}{4\bar{R}} \mathcal L(\bar{R}) \, ,
\end{equation} 
where $ \mathcal L(\bar{R})$ denotes the Lagrangian scalar evaluated on the background.

\paragraph{$f(R)$ gravity.} 
A simple higher curvature gravity is obtained by replacing $R$ in the Einstein-Hilbert action by a function of $R$
\beq
L_{f(R)} = \frac{1}{16\pi G}  \epsilon f(R) \, .
\eeq
To obtain the generalized volume we need
\beq
E^{abcd}_{f(R)} = \frac{f'(R)}{32 \pi G} \left( g^{ac} g^{bd}- g^{ad} g^{bc} \right) \, , \quad E_0 = \frac{f'(\bar{R})}{32 \pi G} \, .
\eeq
The generalized volume then reads
\beq
W_{f(R)} = 
\frac{1}{d-2} \int_{\Sigma} \eta \left [ (d-1) \frac{f'(R)}{f'(\bar{R})} - 1\right] \, .
\eeq
\paragraph{Quadratic gravity.}
A general quadratic   theory of gravity is given by the Lagrangian
\begin{align}\notag
  L_\text{quad}  = & \,   \epsilon \bigg [ \frac{1}{16 \pi G} \big( R -2 \Lambda \big)   + \alpha_1  R^2 + \alpha_2  R_{ab} R^{ab} \\
& + \alpha_3   R_{abcd}R^{abcd}  \bigg ] \, .
\end{align}
Taking the derivative of the Lagrangian with respect to the Riemann tensor leaves us with
\begin{align}\notag
E^{abcd}_\text{quad} = & \, \left ( \frac{1}{32 \pi G} +  \alpha_1 R \right) 2 g^{a[c} g^{d]b} \\
& + \alpha_2 \left (   R^{a[c} g^{d]b} + R^{b[d} g^{c]a} \right) + 2 \alpha_3R^{abcd} \, ,
\end{align}
and using \req{msb} one finds
\beq
E_0 = \frac{1}{32 \pi G} + \left( \alpha_1 + \frac{\alpha_2}{d} +  \frac{2\alpha_3}{d(d-1)} \right) \bar{R}\, .
\eeq
The generalized volume for quadratic gravity thus reads
\begin{align}\notag
&W_\text{quad}
=  \frac{1}{(d-2)E_0} \int_\Sigma \eta \bigg [ (d-1)\left(\frac{1}{32 \pi G}+\alpha_1 R \right) - E_0 \\
&+ \frac{1}{2} \alpha_2 \left (R - R^{ab} u_a u_b (d-2)\right) - 2 \alpha_3 R^{ab} u_a u_b \bigg ].
\end{align}
An interesting instance of quadratic gravity is Gauss-Bonnet theory, which is obtained by restricting to $\alpha_1=-\frac{1}{4}\alpha_2=\alpha_3=\alpha$.
The generalized volume then reduces to
\begin{align} \notag
W_{\text{GB}}  = & \, \frac{1}{(d-2)E_0}\int_\Sigma \eta \left[ \frac{1}{32 \pi G}(d-1) - E_0 \right. \\ \label{eq:WGB}
&\left.+ (d-3) \alpha \Big( R + 2 R^{ab} u_a u_b \Big) \right] \, ,
\end{align}
with $E_0=1/(32 \pi G)+\alpha \bar{R} (d-2)(d-3)/(d(d-1))$.
Since the extrinsic curvature of $\Sigma$ vanishes in the background, 
the structure $R+2R^{ab}u_a u_b$ is equal to the intrinsic Ricci scalar of $\Sigma$,
in the background and at first order in perturbations.


\section{Linearized equations of motion for higher curvature gravity using RNC} \label{app:FLDMRNC}
 
The variational identity (\ref{eqn:offshelllocalgeo}) states that the vanishing of the linearized constraint equations $\delta C_\zeta$ is equivalent to a relation between the variation of the Wald entropy, generalized volume, and matter energy density.
In \cite{Jacobson:2016aa}, Jacobson used this relation to extract  the Einstein equations,
 making use of Riemann normal coordinates. 
Here we perform a similar calculation for the higher curvature generalization of the first law of causal diamond mechanics which will produce the linearized equations of motion.
In this appendix we will restrict to theories whose Lagrangian depends on the metric and the Riemann tensor, $L[g_{ab},R_{abcd}]$, and to linearization around flat space.

The equations of motion for such a general higher curvature theory  read
\begin{equation}\label{eomhighapp}
\begin{split}
  - \frac{1}{2} g^{ab} \mathcal L  + E^{aecd} \tensor{R}{^b_{ecd}} - 2  \nabla_c \nabla_d E^{acdb}    = \frac{1}{2} T^{ab}.
\end{split}
\end{equation}
Linearizing the equations of motion around flat space leads to
\begin{align}\label{lineomhighapp}
 &  - \frac{1}{32\pi G} \eta^{ab} \delta R  + E^{aecd}_\text{Ein} \delta \tensor{R}{^b_{ecd}} - 2  \partial_c \partial_d \delta E^{a c d b}_{\text{higher}} \nonumber\\
&\quad= \frac{\delta G^{ab}}{16\pi G} - 2  \partial_c \partial_d \delta E^{a c d b}_{\text{higher}} = \frac{1}{2} \delta T^{ab}\, ,
\end{align}
where we split $E^{abcd}=E^{abcd}_{\text{Ein}}+E^{abcd}_{\text{higher}}$ into an  Einstein piece, which goes into the Einstein tensor, and a piece coming from higher derivative terms.
We used the fact that many of the expressions in \eqref{lineomhighapp} significantly simplify when evaluated in the Minkowski background because the curvatures vanish.
For example, one might have expected additional terms proportional to the variation of the Christoffel symbols coming from $\delta(\nabla_c\nabla_d E^{acdb})$.
To see why these terms are absent, it is convenient to split this expression into its Einstein part and a part coming from higher derivative terms.
The Einstein piece does not contribute since $E^{acdb}_{\text{Ein}}$ is only a function of the metric and therefore its covariant derivative vanishes.
The higher derivative piece will give $\partial_c \partial_d \delta E^{a c d b}_\text{higher}$ as well as terms such as $\delta \Gamma^{c}_{ce} \nabla_d E^{eadb}_{\text{higher}}$ and $\Gamma^{c}_{ce} \nabla_d \delta E^{eadb}_{\text{higher}}$.
However, the latter two terms are zero because both the Christoffel symbols
 and $E^{eadb}_{\text{higher}}$ vanish when evaluated in the Minkowski background with
 the standard coordinates.

We now want to evaluate each term in \eqref{newvarid} using Riemann normal coordinates.
Taking the stress tensor $T^{ab}$ to be  constant for small enough balls, the variation of \eqref{potati} reduces to
\begin{equation}
\delta H^{m}_{\zeta} = \frac{\Omega_{d-2}\ell^{d}}{d^2-1} \kappa u_a u_b \delta T^{ab} + \mathcal{O}\left(\ell^{d+2}\right)\, ,
\end{equation}
where $\Omega_{d-2}$ denotes the area of the $(d-2)$-sphere, $\ell$ is the radius of our geodesic ball and $u_a$ is the future pointing unit normal.
As was found in \cite{Jacobson:2016aa}, the Einstein piece of the symplectic form will combine with the area term of the entropy to produce the Einstein tensor.
Therefore, we focus on the higher curvature part of $\delta H_\zeta^g$.
Combining \eqref{eqn:hamilton} and \eqref{eqn:symplform}, we find 
{
\makeatletter
\renewenvironment{widetext}{%
  \par\ignorespaces
  \setbox\widetext@top\vbox{%
   \hb@xt@\hsize{%
    \leaders\hrule\hfil
    \vrule\@height6\p@
   }%
  }%
  \onecolumngrid
  \vskip10\p@
  \dimen@\ht\widetext@top\advance\dimen@\dp\widetext@top
  \cleaders\box\widetext@top\vskip\dimen@
  \vskip6\p@
  \prep@math@patch
}{%
  \par
  \vskip6\p@
  \setbox\widetext@bot\vbox{%
   \hb@xt@\hsize{\hfil\box\widetext@bot}%
  }%
  \dimen@\ht\widetext@bot\advance\dimen@\dp\widetext@bot
  \cleaders\box\widetext@bot\vskip\dimen@
  \vskip8.5\p@
  \twocolumngrid\global\@ignoretrue
  \@endpetrue
}%
\makeatother
\begin{widetext}
\begin{align}\label{eq:symplformhigh}
\delta H^g_{\zeta,\text{higher}} 
&= - \frac{4\kappa}{\ell} \int d\Omega \int dr r^{d-2} u_a u_d \eta_{bc } \Big(\delta  E^{abcd}_{\text{higher}}(0)
+ \partial_i \delta  E^{abcd}_{\text{higher}}(0) r n^i + \frac{1}{2}\partial_i \partial_j \delta  E^{abcd}_{\text{higher}}(0) r^2 n^i n^j + \mathcal{O}\left(r^3\right)\Big) \nonumber\\
&= - 4\kappa \Omega_{d-2} \ell^{d-2} u_a u_d \eta_{bc } \bigg(\frac{\delta  E^{abcd}_{\text{higher}}(0)}{(d-1)}
 + \frac{\ell^{2}  \delta^{ij} \partial_i \partial_j \delta  E^{abcd}_{\text{higher}}(0)}{2(d^2-1)} \bigg) + \mathcal{O}\left(\ell^{d+2}\right)\, .
\end{align}
Here, $n^i$ is the normal vector to $\partial \Sigma$ and the indices $a,b$ run over space-time directions, while the indices $i,j$ run only over spatial directions, and $\partial_i$ is the derivative operator compatible with the flat background metric on $\Sigma$.
In the first line, we simply use the formula for the Taylor expansion of a quantity $f$
in the coordinate
system compatible with $\partial_i$, 
\begin{equation}\label{Taylor}
f(x)=f(0)+\partial_a f(0) x^a + \frac{1}{2} \partial_a \partial_b f(0) x^a x^b + \mathcal{O}\left(x^3\right) \, ,
\end{equation}
where $(0)$ denotes that a term is evaluated at $r=0$.
Since we evaluate our expressions on a constant timeslice at $t=0$, we have $x^t=0$ and $x^i=r \, n^i$, where $r$ is a radial coordinate inside the geodesic ball and the index $i$ runs only over the spatial coordinates.
To evaluate the spherical integral, it is useful to note that spherical integrals over odd powers of $n^i$ vanish and furthermore
\begin{align}
\int d\Omega \, n^i n^j &= \frac{\Omega_{d-2}}{d-1}\delta^{ij} \, , \\
\int d\Omega \, n^i n^j n^k n^l &= \frac{\Omega_{d-2}}{d^2-1}\left(\delta^{ij}\delta^{kl}+\delta^{ik}\delta^{jl}+\delta^{il}\delta^{jk}\right)  \;.
\end{align}
Next, we evaluate $\delta S_\text{higher}$, the variation of the higher curvature part of the Wald entropy given in \eqref{eqn:SWald}, in a similar manner.
\begin{equation}
\begin{split}
\delta S_{\text{higher}} = 8 \pi \Omega_{d-2}\ell^{d-2} u_a u_d \left(\frac{\eta_{bc}\delta E^{abcd}_{\text{higher}}(0)}{(d-1)} 
 + \frac{\ell^2\left[\eta_{bc}\delta^{ij} \partial_i \partial_j \delta E^{abcd}_{\text{higher}}(0) 
 	+ 2 \partial_b \partial_c \delta E^{abcd}_{\text{higher}}(0) \right] }{2(d^2-1)}\right)+ \mathcal{O}\left(\ell^{d+2}\right) \; ,
\end{split}
\end{equation}
We are now ready to evaluate the first law of causal diamond mechanics \eqref{newvarid}.
Interestingly, the leading order pieces of the Hamiltonian and Wald entropy exactly cancel against each other.
Note that these two terms would have otherwise dominated over the Einstein piece.
Furthermore, the second term in the symplectic form and Wald entropy also cancel, leaving only a single term from the higher curvature part of the identity.
Including the Einstein piece, we find the first law  for higher curvature gravity reads in Riemann normal coordinates 
\begin{equation}\label{eomRNC}
\begin{split}
- \frac{\kappa \Omega_{d-2}\ell^{d} }{d^2-1} u_a u_d \Big ( \frac{\delta G^{ad}(0)}{8 \pi G} - 4 \partial_b \partial_c \delta E^{abcd}_{\text{higher}}(0) - \delta T^{ad} \Big) 
 +\mathcal{O}\left(\ell^{d+2}\right) =0 \, ,
\end{split}
\end{equation}
proving equivalence to the linearized equations (\ref{lineomhighapp}).
\end{widetext}
}

 
\bibliography{refs-enthigher}{}

\end{document}